\documentclass[
  journal=pasa,
  manuscript=research-paper, 
  year=202X,
  volume=YY,
]{cup-journal}

\usepackage{amsmath}
\usepackage{amssymb,microtype,siunitx,booktabs}
\usepackage{mathrsfs}
\usepackage[normalem]{ulem}   
\usepackage{rotating}
\usepackage{tablefootnote}
\usepackage{xcolor}
\usepackage[breaklinks,colorlinks,urlcolor=blue,citecolor=blue,linkcolor=blue]{hyperref}

\newcommand\degr{\ensuremath{^\circ}}

\newcommand\arcsec{\ensuremath{^{\prime\prime}}}

\newcommand\fdg{\ensuremath{\overset{\circ}{.}}}

\title{First results of sub-arcsecond scale objects identified with ASKAP using interplanetary scintillation}

\author{R.~Chhetri}
\affiliation{CSIRO, Space and Astronomy, P.O. Box 1130, Bentley, WA 6102, Australia}
\email[R.~Chhetri]{rajan.chhetri@csiro.au}

\author{J.~S.~Morgan}
\affiliation{CSIRO, Space and Astronomy, P.O. Box 1130, Bentley, WA 6102, Australia}

\author{R.~D.~Ekers}
\affiliation{CSIRO Space and Astronomy, P.O. Box 76, Epping, NSW 1710, Australia}

\author{E.M.~Sadler}
\affiliation{Sydney Institute for Astronomy, School of Physics A28, The University of Sydney, NSW 2006, Australia}
\alsoaffiliation{CSIRO, Space and Astronomy, PO Box 76, Epping, NSW 1710, Australia}

\author{V. A. ~Moss}
\affiliation{CSIRO Space and Astronomy, P.O. Box 76, Epping, NSW 1710, Australia}

\author{A.~Waszewski}
\affiliation{School of Physics, The University of Newcastle, Callagham, NSW 2308, Australia}
\alsoaffiliation{CSIRO, Space and Astronomy, P.O. Box 1130, Bentley, WA 6102, Australia}

\author{Paris E. ~Gordon-Hall}
\affiliation{Sydney Institute for Astronomy, School of Physics A28, The University of Sydney, NSW 2006, Australia}
\alsoaffiliation{CSIRO, Space and Astronomy, PO Box 76, Epping, NSW 1710, Australia}

\received {dd Mmm YYYY}
\revised  {dd Mmm YYYY}
\accepted {dd Mmm YYYY}
\published{22 September 202X}


\begin{document}

\begin{abstract}
We present a catalogue of 131 compact ($\lesssim0.1\arcsec$) sources detected at 823\,MHz via their Interplanetary Scintillation (IPS).
These measurements were made with the ASKAP telescope, across its full field of view of 35 square degrees, utilising all 36 Phased Array Feed (PAF) beams. To bypass ASKAP's standard correlator's minimum integration limit of 10\,s, we used the CRAFT data capture system (CRACO), with visibilities sampled every 110\,ms.  Here we present the data processing steps, the sources detected, and their IPS-inferred properties. ASKAP IPS cleanly separates two populations: compact hot spots embedded in extended lobes and IPS-unresolved sources which are AGN or CSO sources associated with the galactic nucleus. We also compare these results with the results from observations of IPS at 162\,MHz with the Murchison Widefield Array, providing the spectra of compact components between 162\,MHz and 888\,MHz. These measurements further re-enforce the dominance of peaked-spectrum SEDs in the compact-source population at frequencies below 1\,GHz. This pilot study using test data is a pathfinder for a more comprehensive ASKAP IPS survey which is underway.
\end{abstract}

\section{Introduction}

Development of widefield instruments at frequencies below 1\,GHz, namely the Australian SKA Pathfinder \citep[ASKAP;][]{Johnston2007,Johnston2008,Hotan2021}, the Murchison Widefield Array \citep[MWA;][]{Tingay2013}, Low Frequency Array \citep[LOFAR;][]{vanHaarlem2013} have provided excellent deep surveys of the sky with high quality spectral information for radio sources detected at these lower frequencies (e.g. GLEAM \citealp{HurleyWalker2017}; LoTSS \citealp{Shimwell2017}; TGSS \citealp{Intema2017}).

However, Low-frequency all-sky surveys lack the sub-arcsecond resolution required to probe the kpc-scale properties of the detected sources
 \citep[1\,kpc$\sim$0.2\arcsec\ at $z$$\sim$1; the median redshift of sources in millijansky-level surveys, see e.g.][]{brookes2008}.
This is a physically-interesting size range for the study of young and recently-triggered radio sources, including the peaked-spectrum (PS) and compact steep-spectrum (CSS) radio sources \citep{ODea1998}, which have been identified in large numbers at low frequencies \citep{Callingham2017}, as well as the rarer Compact Symmetric Objects \citep[CSOs;][]{Readhead2024}. 

In the Northern Hemisphere, wide-field imaging with international baselines of LOFAR \citep[e.g.][]{Sweijen2022} provides complementary information on the sub-arcsecond structures of the low-frequency (100-200\,MHz) radio source population, albeit with the huge computation costs associated with gigapixel interferometric imaging.

Measurements with Interplanetary Scintillation (IPS), provide a much simpler pathway to obtain information on these critical sub-arcsecond scales.
These random intensity fluctuations, induced by the solar wind plasma \citep{Clarke1964, Hewish1964}, are only coherent for sufficiently small sources, which means that IPS variability is strongly suppressed for sources of size $\gtrsim$1\arcsec.

\citet{Morgan2018-IPS1} demonstrated that interferometric imaging of individual snapshots provides an efficient way to identify and characterise the IPS of large number of sources for widefield instruments, and detected IPS from hundreds of sources across the MWA field of view in a 5-minute test observation.
\citet[][]{Chhetri2018_IPS2} demonstrated that the Normalised Scintillation Index (NSI) can be used to quantify IPS, where an NSI$\sim$1 source scintillates like a point source, implying that the source is small compared to the Fresnel angle, $\theta_F$, where $\theta_F/\mathrm{arcsec} = 0.3\sqrt{\nu/150\mathrm{MHz}}$ for IPS \citep[][Equation~4]{Chhetri2018_IPS2}.
The NSI is therefore a measure of the fraction of flux which is compact relative to the Fresnel angle. 

The high efficiency of IPS means that surveys covering a large fraction of the accessible sky have already been carried out.
The first results from the MWA Phase II IPS survey \citep{Morgan2022_IPS_DR1} represents the largest single data release of compact radio sources ever, and the full accessible sky has now been surveyed with the MWA (Waszewski et al. in prep.).
\citet{Jackson2022} also show that there is good agreement between the MWA IPS (162\,MHz) and LOFAR VLBI (140\,MHz) estimates of source size for objects in an overlapping region of sky.  

\citet{Chhetri2018-IPS3,Chhetri2018_IPS2} presented an overview of the properties of the compact radio source population at 162\,MHz as revealed by IPS.
They showed that in contrast to the population observed above 1\,GHz, peaked-spectrum sources dominate.
Furthermore, blazar objects are often revealed by IPS to be embedded in more extended structure which is more prominent at lower freqencies, implying that selection based on surveys at GHz frequencies  is likely to miss a significant population of low-frequency compact sources. 

This provides an important motivation for making IPS measurements with ASKAP: the low-frequency compact source population is quite distinct from the picture that has developed at 1\,GHz and above \citep[e.g.][]{Chhetri2013-AT20G_HARC}, and so ASKAP IPS provides a useful intermediate measurement: both in frequency (at 823\,MHz), and in resolution (0.1\arcsec\ vs 0.3\arcsec for MWA IPS and $\sim$0.01\arcsec for L-band VLBI).
Additionally, there is evidence from the First Large Absorption Survey in HI \citep[FLASH;][]{Allison2022} that HI absorption is more detectable in radio sources with compact structures, which can be identified using IPS in combination with the FLASH survey \citep[e.g.][]{Aditya2024}.
Wide-field IPS surveys also provide an efficient way of identifying a grid of candidate phase calibrators for SKA at frequencies below 1\,GHz, just as compact sources from the 20\,GHz AT20G survey \citep{murphy2010,Chhetri2013-AT20G_HARC} provided essential input to the calibrator catalogue for ALMA \citep{petrov2011}. 
Finally, wide-field IPS observations with a high density of sight lines provide a powerful tool for studying the solar wind and space weather \citep{Morgan2023,Waszeski2023_paper1,Waszeski2025_paper2,Waszeski2025_paper3}.

ASKAP is an excellent instrument for this work, due to its wide field of view (and sufficient instantaneous UV coverage for high-fidelity imaging), and the high time resolution provided by the CRAFT COherent imaging (CRACO) system \citep{Wang2025}.
Additionally, its autonomous operations model makes it easy to schedule observations according to the constraints of IPS observations.
Previously, using only 3\,s of data, the possibility of making IPS observations with ASKAP has been demonstrated \citep{Chhetri2023_IPS-ASKAP}, using a predecessor to CRACO. 

In order to test the technical feasibility of performing IPS observations with ASKAP, a series of test observations were conducted in early Feb 2023. In this paper, we present the results of one of these observations: first results from IPS observations using all 36 phased array feed (PAF) beams of ASKAP. We show that ASKAP can be used to conduct a survey for sub-arcsecond scale radio sources across large areas in the sky and discuss the properties of compact objects detected at 823\,MHz. 

The layout of this paper is as follows: in Section 2, we present the data acquisition and processing. In Section 3, we present our results and discussion. We present our conclusions in Section 4. We also include a catalogue of detected compact objects and upper limits for non-detections with this paper. We use the convention of positive index for relationship between flux density ($S$), frequency ($\nu$) and spectral index ($\alpha$) as: 
\mbox{$S_{\nu}\propto\nu^{\alpha}$}.

\section{Data acquisition and processing}
\subsection{The ASKAP telescope}
\label{subsec:askap}
The ASKAP telescope is a wide field-of-view instrument located at the Inyarrimanha Ilgari Bundara; the CSIRO Murchison Radio-astronomy Observatory in Western Australia \citep{Hotan2021}. It has 36$\times$12\,m antennas in a 2D array extending to a maximum spacing of 6\,km. The extremely wide  ($\sim$30\,deg$^{2}$) Field of View (FoV) of ASKAP is achieved by the use of a Phased Array Feed (PAF) in the focal plane of each telescope which produces images for 36 beams simultaneously, as illustrated in Figure~\ref{Fig:ASKAP_obs}. In addition, the telescope is equipped with a system for high-time resolution imaging, which is discussed in detail in Section~\ref{subsec:craco}. These two features make ASKAP highly capable of IPS observations.

In addition to these unique hardware features, ASKAP has also adopted an autonomous science operations model since 2020, based on the SAURON scheduler and increased system automation (Scheduling Autonomously Under Reactive Observational Needs, Moss et al. in prep). This model means that the addition of new science cases with specific constraints and requirements (such as the IPS observations described here) can be more readily incorporated alongside existing ASKAP observations, as long as these new cases maximise their flexibility in order to avoid disruption.

\subsection{High time resolution data from ASKAP using CRACO}
\label{subsec:craco}
To overcome the limitation of the 10\,s time resolution of the standard ASKAP correlator we used a commensal backend system on ASKAP developed by the Commensal Real-Time ASKAP Fast Transient \citep[CRAFT,][]{Macquart2010} team. 
The CRAFT COherent imaging system ``CRACO'' is designed for detection of Fast Radio Bursts, and the reader is referred to \citet{Wang2025} and references therein for a full description of CRACO and its capabilities. For IPS observations, the full CRACO system is not utilised directly. Instead we use a specific capability to dump visibilities at 110\,ms time resolution, which (as we will show) is an optimal cadence for fully resolving IPS variability at our observing frequency.

The use of CRACO to produce this data product is currently undergoing commissioning as part of a transition to National Facility status, and the data presented in this work was collected as part of CRACO commissioning tests. Hardware limitations on CRACO at the time meant that the only 128\,MHz (out of total 288\,MHz) of bandwidth was possible and data from only the inner 30 ASKAP antennas (out of the total 36 antennas) could be collected.  CRACO combines the XX and YY polarisation channels only outputting single channel data after integration. The observing bandwidth was placed at the bottom end of ASKAP band 1, and was divided into 768 channels (each 166.7\,kHz wide), thereby covering 759.5\,MHz--887.5\,MHz (midpoint frequency 823.5\,MHz). Each of the 16 observations made for our test was 2.5 minutes long. The visibilities for each PAF beam were 4.2\,GB in size, with the total of 217\,GB for each observation after imaging. Table~\ref{Table:obs_details} summarises these details of our observation. 

\begin{figure}[hbt!]
	\centering
	\includegraphics[width=0.91\linewidth]{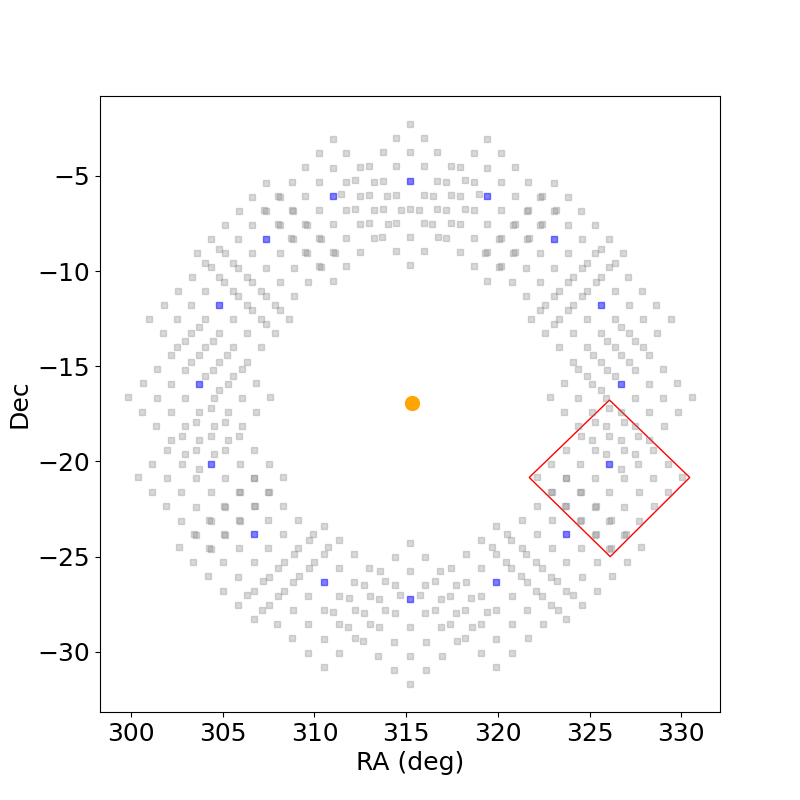}
	\caption{The 16 observations made on 2 Feb 2023 around the Sun (marked with yellow circle) with ASKAP. The pointing centre of the central PAF beam (beam 00) is shown in blue colour, and the other pointing centres of other beams are shown in grey. This paper is based on the single observation highlighted by the red square in the plot.}
	\label{Fig:ASKAP_obs}
\end{figure}

The ASKAP telescope was used to make measurements of IPS on 2 Feb 2023, with 16 $\times$ 2.5\,min observations around the Sun made at the solar elongation of $\sim$10\degr\ at the centre of the PAF footprint, and covering solar elongations between 7\fdg 5 and 12\fdg5 (Figure~\ref{Fig:ASKAP_obs}). To match the scientific requirements for IPS investigation, we worked with the ASKAP operations team to define a new mode for SAURON that would collect data under the right constraints for both the target fields and bandpass calibrator. In this paper we present the results from one of the 16 scheduled blocks marked with a red square in Figure~\ref{Fig:ASKAP_obs}, (SBID 47529, pointed at J2000.0 Right Ascension = 21:44:20.37 Declination = -20:53:18.44).

The output visibilities from CRACO were converted to MIRIAD format for calibration and imaging with the MIRIAD software \citep{1995ASPC...77..433S}. All data processing was done using the CETUS server (32 cores, 504 GB memory) at Australia Telescope National Facility (ATNF), with each PAF beam processed in $\sim$5 hours (from pre-processing, calibration to imaging).

\subsection{Calibration and imaging}
\label{Sec:Cal+Img}
We used observations of PKS 1934-638 for our bandpass and gain calibrations. PKS 1934-638 is a strong compact source which is used as the primary calibrator for the Australia Telescope Compact Array (ATCA).
The calibration solutions were obtained for each PAF beam using observation of PKS 1934-638 centred in the beam. These calibration solutions were applied to the target field data observed using the corresponding PAF beam. 

\begin{figure}[hbt!]
\centering
\includegraphics[width=1.0
\linewidth]{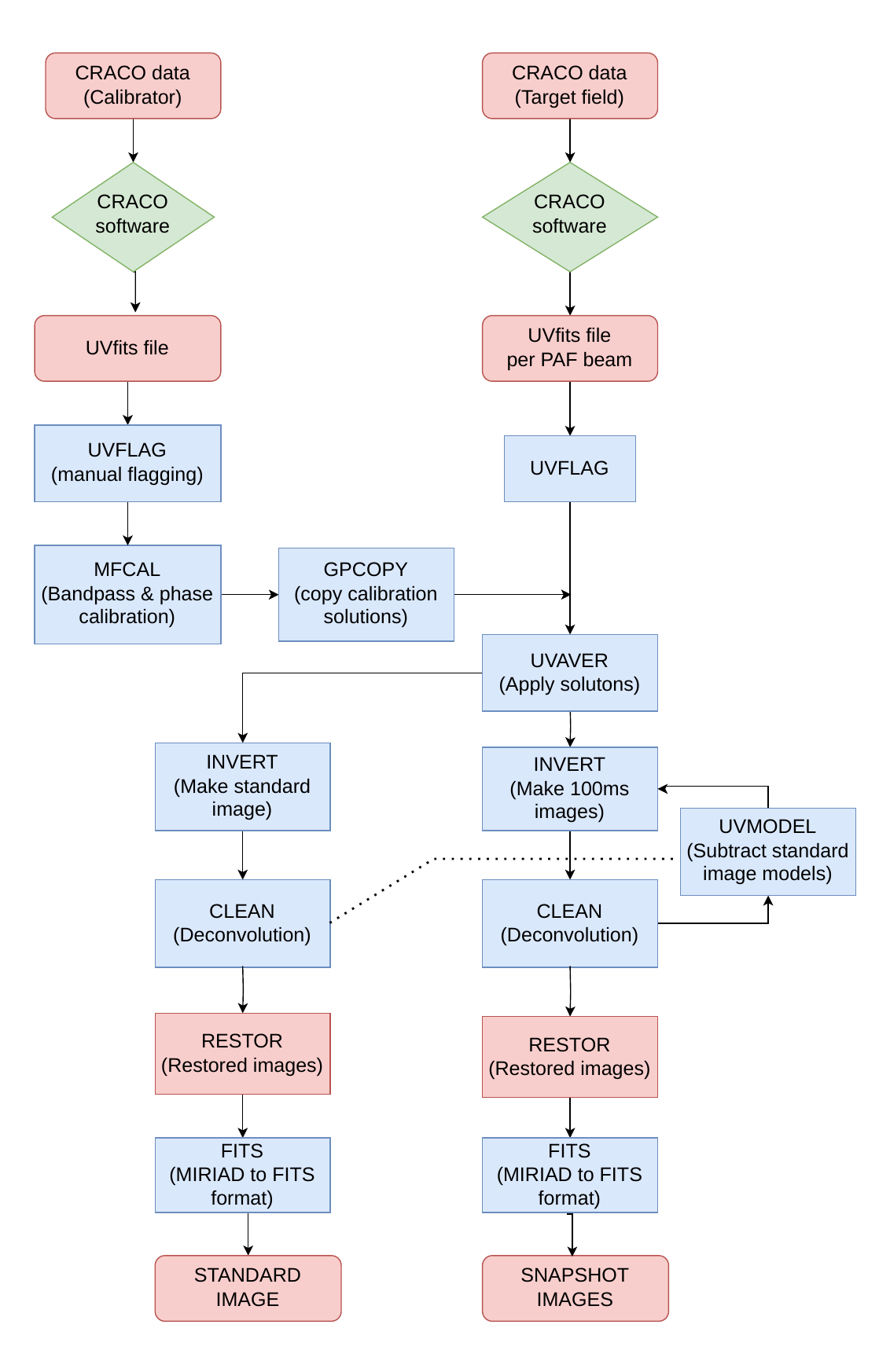}
\caption{The flow chart shows the steps used in making our images. Pink coloured boxes indicate data products, green coloured boxes indicate CRACO specific steps and blue coloured boxes indicate MIRIAD specific steps. The dotted line indicates that the source model was obtained from the standard image (on the left) and used to subtract from the 100\,ms images (on the right of the flow chart).}
\label{fig_sim}
\end{figure}

We set up a custom pipeline to process the data using the MIRIAD software (summarised in Figure~\ref{fig_sim}). Essentially, the pipeline produces for each PAF beam separately:
(1) images made at 110\,ms intervals for the duration of the observation, and
(2) a single image for the total duration of the observation. Following our nomenclature from MWA IPS observations we term these the ``snapshot images'' and the ``standard image'' respectively. The output snapshot images are stored in an HDF5 container, ordered such that time series for a given pixel can easily be extracted \citep[][Appendix~1]{Morgan2018-IPS1}. 

\begin{table}[hbt!]
		\caption{Details of observation and acquired data for one target field}
		\label{Table:obs_details}
		\begin{tabular}{ll}
			\toprule
			Details & Values\\
			\hline
			
			Central Frequency (MHz)\tablefootnote{Only data from the lower half of ASKAP frequency band were obtained} & 823.5\\ 
			Bandwidth (MHz) & 128\\
			Polarisation & Total intensity (XX+YY) \\
			No. antennas\tablefootnote{Only the inner 30 antennas are currently used for CRACO observations, out of which 3 were not functioning during our observations.} & 27 \\
			No. of PAF beams & 36 \\
			Footprint  name& square$\_$6$\times$6 \\
			Footprint pitch & 1.05 \\
			Footprint rotation (deg) & 45 \\
			Correlator integration time (ms) & 110.592\\
			Total integration per target field (sec) & 150\\
			Scan date and time (UTC) & 2023-02-02 05:17:08 \\
			Calibrator field & PKS 1934-638 \\
			\hline
			
			\bottomrule
		\end{tabular}

\end{table}


Baselines shorter than 85\,k$\lambda$ ($\sim 310$\,m) were flagged to avoid the adverse effect in the image due to the Sun (which is mostly resolved out by longer baselines) being in the far sidelobes of the primary beam.

All our images are 1250$\times$1250 pixels wide, with each pixel 10\arcsec$\times$10\arcsec in size. To deconvolve (clean) our images we first made an initial image  using total observing time and applied minimal cleaning. We used the source finding algorithm Aegean \citep{Hancock2018} on this image to detect sources with signal-to-noise ratio $>$5 and defined a $6\times6$ pixel wide clean box around each component that was detected. We then restarted deconvolution using these clean boxes to obtain our standard images. 

\subsection{Variability images from snapshots}
Each 110\,ms snapshot was then imaged. To reduce the computation required, and avoid problems associated with non-linear CLEAN effects, we subtracted the clean component models obtained for our standard image from the visibilities using the MIRIAD task UVMODEL. This mean-subtracted visibility data was used to produce all our snapshot images\footnote{We note here that constraining clean areas in the manner stated above was important to avoid unconstrained cleaning which manifested as variability in non-scintillating sources in the variability image.}.  

We produced a ``variability image'' for each PAF beam by taking a standard deviation for each pixel in the snapshot images along the time axis \citep[see][for details of this step]{Morgan2018-IPS1}. Before taking the standard deviation, the timeseries is filtered in the time-domain by 2nd-order butterworth filters with a high-pass and low-pass filter frequencies  of 0.07\,Hz and 1.219\,Hz respectively. This filter emphasises IPS frequencies and removes any spurious red noise due to (e.g.) ionospheric scintillation. The result is an image with a constant offset from zero (due to thermal noise) with IPS sources sitting above this noise floor. This image has Gaussian noise statistics, meaning that standard source-finding tools can be used to identify scintillating sources, as detailed in the following section.

\subsection{Identifying scintillating sources}

\begin{figure}[hbt!]
	\centering
	\includegraphics[width=1.0\linewidth]{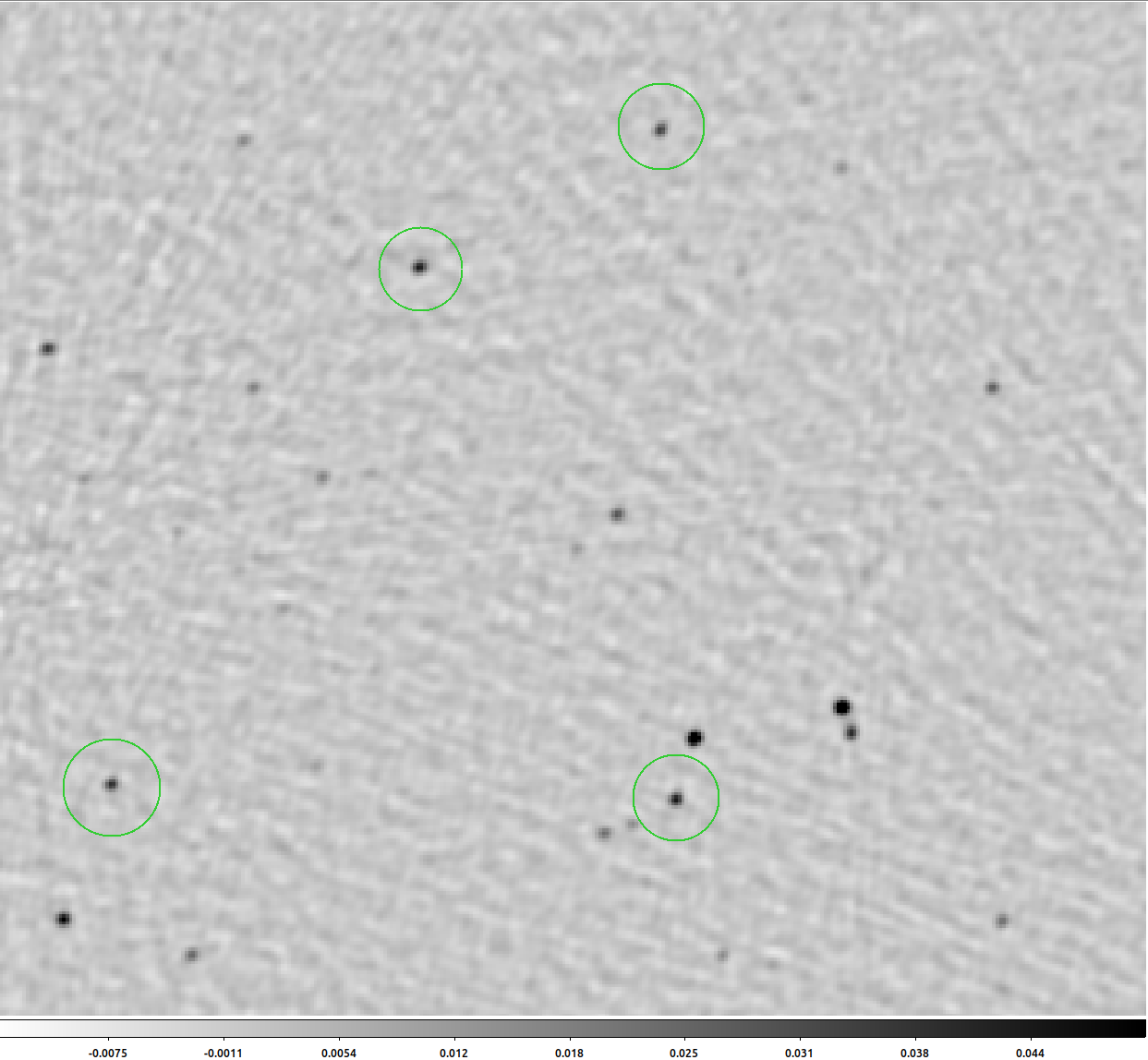}
	\includegraphics[width=1.0\linewidth]{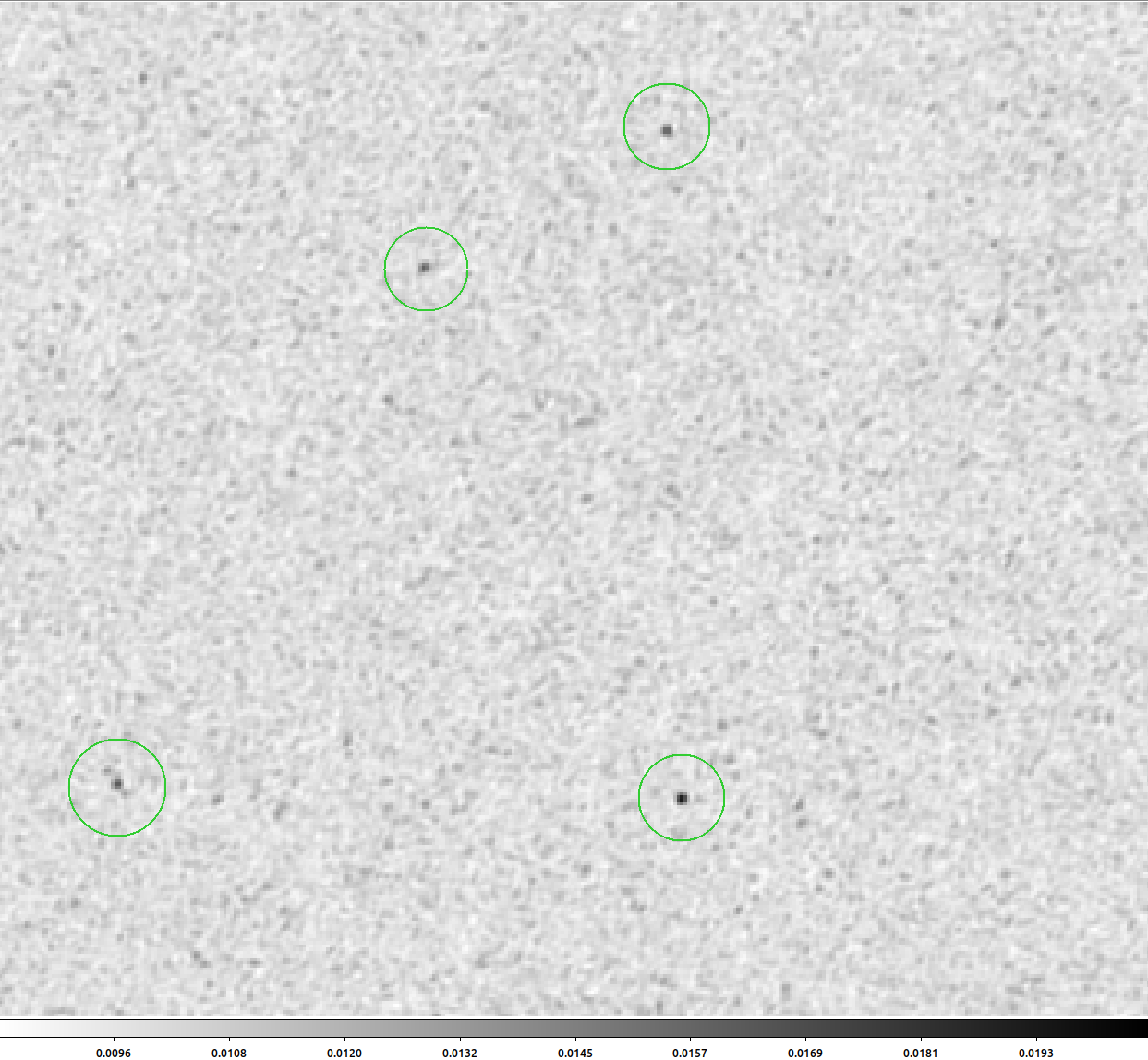}
	\caption{Example of the inner part of the standard image for one beam (beam 00), and a corresponding variability image in the same area (top and bottom respectively). The images are $\sim$1\degr\ across. Note the change in the density of sources. Sources showing high level of scintillation are identified with green circles.  }
	\label{Fig:img_beam00_cont+var}
\end{figure}

As discussed above, we processed observations from different PAF beams separately; however to illustrate the process used, we will focus on one PAF beam. The other 35 PAF beams are all processed identically (and can be processed in parallel). Figure~\ref{Fig:img_beam00_cont+var} shows an example of the quality of image obtained for our observation for beam 00. The density of sources in the variability image is significantly lower, indicating that not all sources detected in the standard image have compact components above our detection threshold. 

\begin{figure*}[hbt!]
	\centering
	\includegraphics[width=1.0\linewidth]{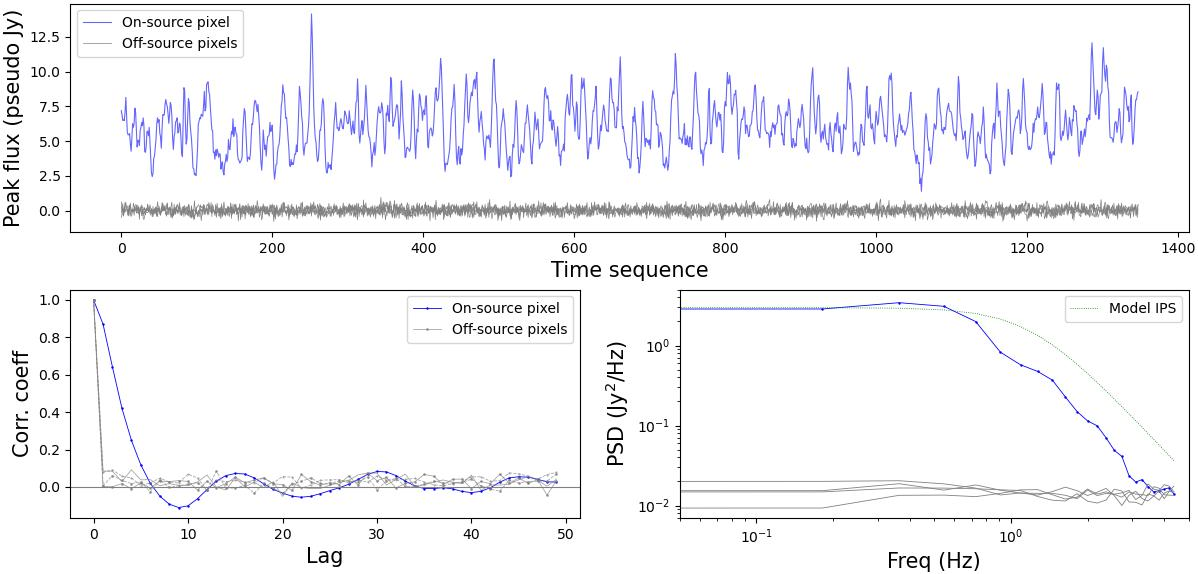}
	\caption{Time series (top panel) and autocorrelation function (bottom left panel) and power spectrum (bottom right panel) of a bright strongly scintillating source identified in our observation. The blue lines indicate the data for the observed source, and the grey lines indicate variability due to noise only from positions offset by 25 pixels away from the source in top, bottom, left and right directions. `Time sequence' in the top plot X-axis indicates the number of 110\,ms time intervals since the beginning of observation, and `Lag' in bottom left plot X-axis indicates number of lag intervals. Modelled IPS power spectrum for a point-like object is also plotted as a reference in the bottom right plot with green dotted line. }
	\label{Fig:ts+ACF}
\end{figure*}

We present a timeseries for a strongly scintillating source along with its autocorrelation function and power spectrum in Figure~\ref{Fig:ts+ACF}. For clarity, time series for only the first 12\,s of data is shown while our total observation is 2.5 minutes long. To demonstrate that variability induced due to noise only is distinct from the variability due to IPS for the presented source we also plot the time series and autocorrelation function for pixels offset by 25 pixels from the position of the source. 

In order to identify scintillating sources, we ran the source finding algorithm Aegean independently (with the default detection cut-off of 5$\sigma$) on the standard and variability images. We imposed the condition that each detection in the variability image must have a corresponding detection in standard image (within 15\arcsec). For these detections, we estimated the scintillation index using scintillating flux density obtained from the variability image. This requires subtracting the noise in quadrature. For detailed description of the determination of the scintillating flux from the variability image, the reader is referred to Sections~2.3 and 3.2 in \citet{Morgan2018-IPS1}.

\begin{figure}[hbt!]
	\centering
	\includegraphics[width=1.0\linewidth]{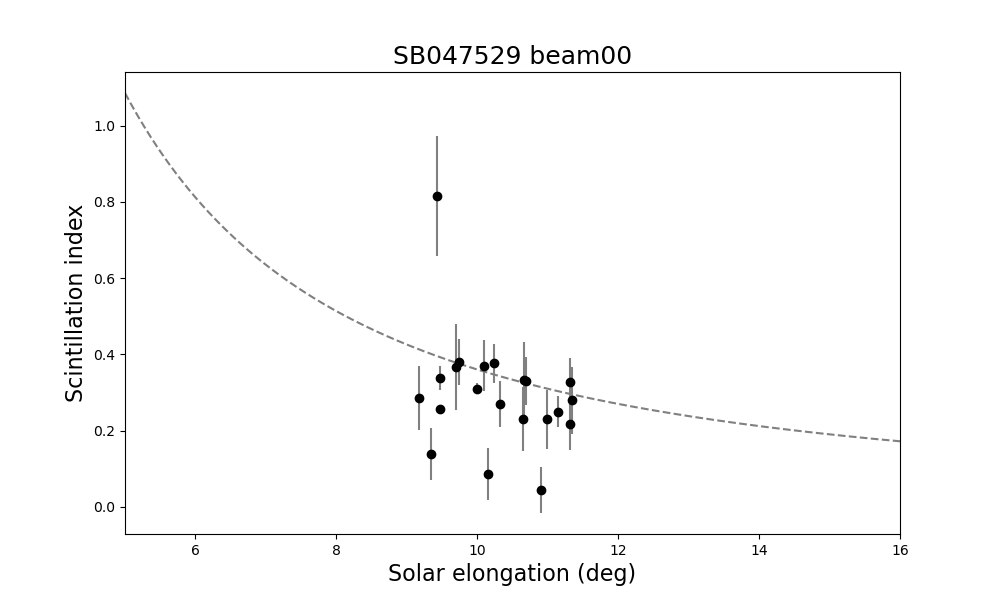}

	\caption{Scintillation index as a function of solar elongation for detected sources in beam 00. The dotted line is an empirical estimate of scintillation expected for a point-like object from \cite{Rickett1973}. The outlier is not a real source but a detection due to a noise on the outer part of the beam as discussed in the text.}
	\label{Fig:solEl-SI_beam00}
\end{figure}

The scintillation index is then obtained as:
\begin{equation}
	SI = \frac{\Delta S_\mathrm{scint}}{S}
\end{equation}
where $S$ is the flux density obtained from the standard image.
Figure~\ref{Fig:solEl-SI_beam00} shows the scintillation index as a function of solar elongation obtained from this step, compared to the maximum scintillation index expected of a point source ($m_{pt}$) from \cite{Rickett1973}.

We then calculated the Normalised Scintillation Index \citep[NSI, defined in][]{Chhetri2018_IPS2} for each source to remove the effects of 
different solar elongations, so that direct comparisons of sources was possible, as:
\begin{equation}
	NSI = \frac{SI}{m_{pt}}
\end{equation}

NSI provides a measure of angular size: sources with NSI = 1 correspond to sources equal to or smaller than the Fresnel size ($\sim$0.1\arcsec\ at our observed frequency). Figure 5 in  \cite{Morgan2019_IPS5}, and the associated text, provide a detailed description the angular size constraints provided by NSI. 

In addition to angular size, NSI provides a measure of the fraction of total flux density that arises from the compact component. Therefore, by having estimates of NSI at different frequencies, it is possible to obtain the spectral energy distribution of the compact component even when embedded within a larger structure. 
We discuss this opportunity further in Section~\ref{Sec:Results}.

\subsubsection{Estimates of upper limits}

We estimated upper limits of NSI for all sources detected in the standard image but undetected in the variability image. For this we used the background and RMS values in the variability image at the position of each source, obtained from the Background And Noise Estimator \citep[BANE; ][]{Hancock2018} run on the variability image. To estimate this upper limit, we used a 5-$\sigma$ value (thereby ensuring that all continuum-detected sources either have a detection or a detection limit). These values can be converted to an upper limit on $\Delta S_\mathrm{scint}$ via the \citet{Chhetri2018_IPS2} Equation~2, which in turn can be converted to an NSI upper limit via equations~1\&2 above.

\subsection{Combining measurements from different PAF beams into a single catalogue}
As described in Section~\ref{subsec:askap}, ASKAP achieves its wide field of view by forming 36 beams on the sky simultaneously. For the purposes of the analysis in this paper, these are equivalent to 36 separate observations of neighbouring fields with an inteferometer using conventional single-pixel feeds (in the future we may be able to exploit the fact that these are simultaneous observations, so IPS signatures are in common while the noise of different pointings is at least semi-independent; for now we defer this to future work).

This means that steps outlined in the above section have been carried out independently for each PAF beam. The ASKAP PAF beam weights  \citep{Hotan2021} used for this observation provide some overlapping coverage for areas of the sky observed in more than one PAF beams, so many sources have multiple simultaneous observations. The next task is to combine these separate catalogues into a single catalogue with a single measurement per source.

\begin{figure}[hbt!]
	\centering
	\includegraphics[width=0.95\linewidth]{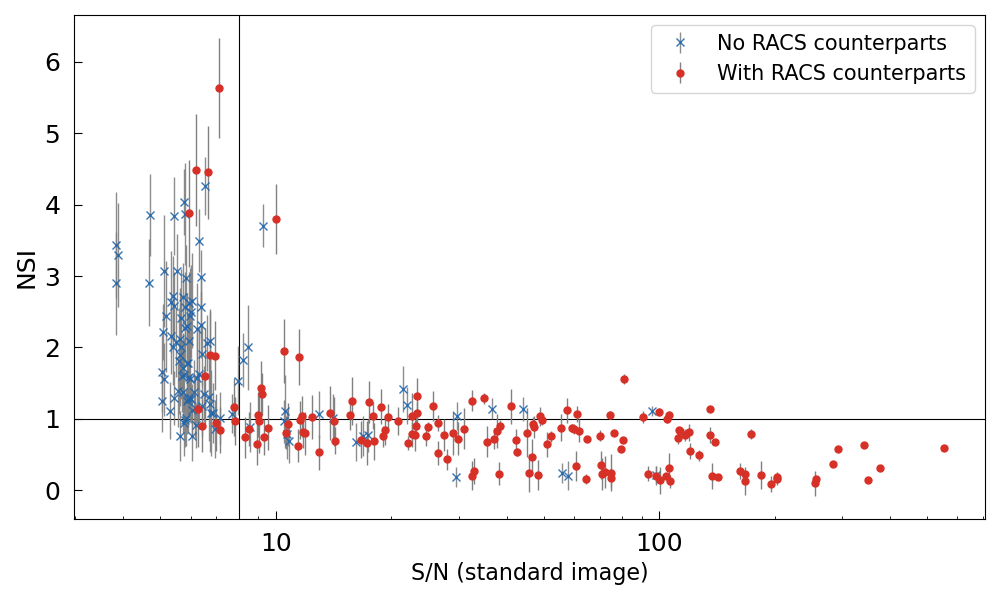}
	\caption{Figure shows NSI as a function of their signal-to-noise (S/N) ratio in the standard images. We combined S/N cut-off of 8 (vertical line) with the requirement that the source has to have a counterpart in RACS-low (shown with red circles) to reduce noise induced spurious detections (shown with blue crosses).}
	\label{Fig:SNcont_NSI_COMBINED}
\end{figure}

There were a total of 528 detections in our images from different PAF beams. To identify true source detections from noise-induced detections, we implemented the constraints that: 
(1) each detection (in the standard image) must have a counterpart present within 15\arcsec\ in the more sensitive Rapid ASKAP Continuum Survey catalogue \citep[RACS-Low catalogue]{Hale2021} made at 888\,MHz, 
(2) that objects must have signal-to-noise ratio of 8 or above in the standard image. The S/N of 8 was used to balance reliability and completeness. 
These constraints significantly reduced the number of noise-induced spurious scintillating detections (as seen in Figure \ref{Fig:SNcont_NSI_COMBINED}). 
411 detections remained (including multiple observations of the same source in different PAF beams) after removing spurious detections.

In order to obtain a representative NSI for each object, we took the measurement where the signal-to-noise ratio of detection in the standard image is the highest when seen in different beams, essentially selecting the detection from the beam with the best quality image and closest to the primary beam centre. We plot these NSI as a function of signal-to-noise in the standard image in Figure~\ref{Fig:SNcont_NSI_COMBINED}. We visually inspected the images for the small number of unusually high NSI detections in Figure~\ref{Fig:SNcont_NSI_COMBINED}. If any detection was found not to be detected in other PAF beam images (when S/N consideration would require them to be detected), such detections were removed as noise peaks. 

\section{Results and discussion}
\label{Sec:Results}

\begin{figure}[hbt!]
	\centering
	\includegraphics[width=0.9\linewidth]{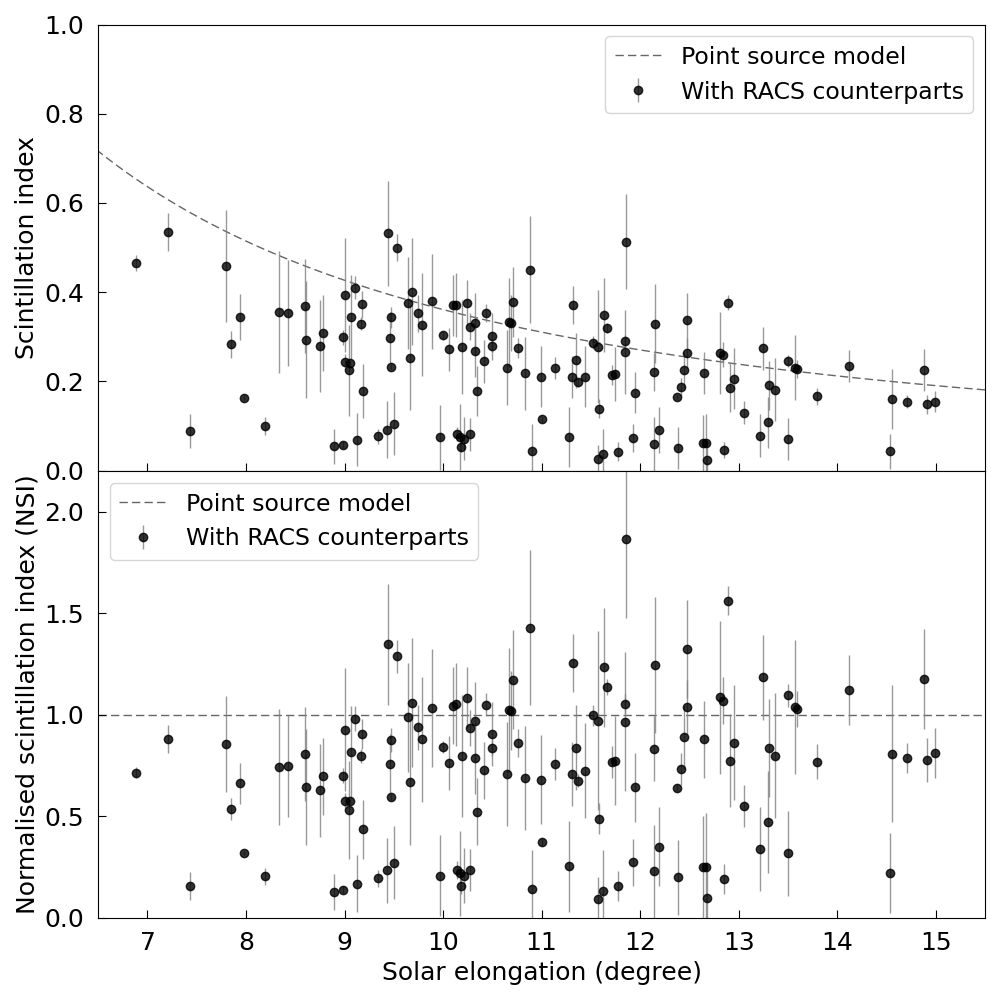}

	\caption{Scintillation index (top panel) and normalised scintillation index (bottom panel) as a function of solar elongation from different PAF beams. The dotted line is an empirical estimate of scintillation expected for a point-like object from \cite{Rickett1973}.}
	\label{Fig:solEl_SI+NSI-COMBINED_v2}
\end{figure}

The analysis outlined in the previous section resulted in a table of 131 scintillating sources,  corresponding to a source density of approximately 3.8 scintillating sources per square degree (using a footprint area of 35 sq.~deg.). The distribution of their scintillation indices and NSI, as a function of solar elongation is presented in Figure~\ref{Fig:solEl_SI+NSI-COMBINED_v2}.
We present our final catalogue in Table \ref{Table:catalogue} (only the first 20 rows are shown in the printed version).

\begin{sidewaystable}[hbt!]
	\scriptsize	
	\setlength{\tabcolsep}{3pt}
	\caption{First 20 rows of the catalogue presented with this paper. Here,
		Column 1: Name of the source in RACS-low (888\,MHz) catalogue, 
		Column 2: Right Ascension (J2000) in RACS-low catalogue, 		
		Column 3: Declination (J2000) in RACS-low catalogue,
		Column 4: Peak flux density of the source presented in RACS-low catalogue, 
		Column 5: Uncertainty in peak flux density, 
		Column 6: Signal to noise ratio in our \textit{standard images}.
		Column 7: Signal to noise ratio in our variability images,
		Column 8: Solar elongation of the source at the time of observation,
		Column 9: ASKAP Normalised scintillation index (NSI) at 823\,MHz,
		Column 10: Uncertainty in ASKAP NSI,
		Column 11: Upper limit for ASKAP NSI for sources not detected in any variability images,
		Column 9: ASKAP Normalised scintillation index (NSI) at 823\,MHz,
		Column 10: Uncertainty in ASKAP NSI,
		Column 11: Upper limit for ASKAP NSI for sources not detected in any variability images,
		Column 12: GLEAM name for the source,
		Column 13: Peak flux density at 162\,MHz, based on the GLEAM survey,  
		Column 14: MWA NSI at 162\,MHz,
		Column 15: Uncertainty in MWA NSI at 162\,MHz,
		Column 16: Upper limit in MWA NSI at 162\,MHz,
		Column 14: MWA NSI at 162\,MHz,
		Column 15: Uncertainty in MWA NSI at 162\,MHz,
		Column 16: Upper limit in MWA NSI at 162\,MHz,
		Column 17: Spectral index of the compact component between 162 and 823\,MHz}
	
	\label{Table:catalogue}	
	\begin{tabular}{|l|l|l|r|r|r|r|r|r|r|r|r|r|r|r|r|c|}
		\hline
		
		\multicolumn{1}{|c|}{RACS-low\_name} &
		\multicolumn{1}{c|}{RA} &
		\multicolumn{1}{c|}{Dec} &
		\multicolumn{1}{c|}{S\_888} &
		\multicolumn{1}{c|}{err\_S\_888} &
		\multicolumn{1}{c|}{SN\_std} &
		\multicolumn{1}{c|}{SN\_var} &
		\multicolumn{1}{c|}{Sol\_el} &
		\multicolumn{1}{c|}{NSI\_ASKAP} &
		\multicolumn{1}{c|}{err\_NSI\_ASKAP} &
		\multicolumn{1}{c|}{ul\_NSI\_ASKAP} &
		\multicolumn{1}{c|}{GLEAM\_name} &
		\multicolumn{1}{c|}{S\_162} &
		\multicolumn{1}{c|}{NSI\_162} &
		\multicolumn{1}{c|}{err\_NSI\_162} &
		\multicolumn{1}{c|}{ul\_NSI\_162} &
		\multicolumn{1}{c|}{alpha\_compact} \\

			& (deg) & (deg) & (mJy/beam) & (mJy/beam) & & & (deg) & & & & (mJy/beam)& & & & & \\
			\textit{[1]} & \textit{[2]} & \textit{[3]} & \textit{[4]} & \textit{[5]} & \textit{[6]} & \textit{[7]} & \textit{[8]} & \textit{[9]} & \textit{[10]} & \textit{[11]} & \textit{[12]} & \textit{[13]} & \textit{[14]} & \textit{[15]} & \textit{[16]} & \textit{[17]}\\
		
		\hline

		J212424.0-204706 & 321.100242 & -20.785213 & 37.4 & 0.4 & 5.8 &  & 6.7 &  &  & 0.50 &   J212423-204710 & 180 &  &  & 0.00 &   \\
		J212432.1-203836 & 321.133754 & -20.643433 & 116.8 & 0.4 & 12.6 &  & 6.7 &  &  & 0.25 &   J212432-203834 & 476.6 &  &  & 0.09 &   \\
		J212545.7-211406 & 321.440688 & -21.235206 & 40.7 & 0.3 & 7.4 &  & 7.2 &  &  & 0.40 &   J212545-211407 & 135 &  &  & 0.40 &   \\
		J212624.8-214943 & 321.603659 & -21.82869 & 164 & 0.3 & 22.5 &  & 7.7 &  &  & 0.18 &   J212624-214942 & 840.9 & 0.32 & 0.02 &  &   \\
		J212641.5-220728 & 321.672972 & -22.1246 & 44.8 & 0.4 & 5.5 &  & 7.9 &  &  & 0.87 &   J212641-220726 & 246.9 & 0.68 & 0.05 &  &   \\
		J212648.4-202025 & 321.701889 & -20.34052 & 59.7 & 0.3 & 8.5 &  & 7.0 &  &  & 0.27 &   J212648-202024 & 237.3 &  &  & 0.25 &   \\
		J212658.4-195715 & 321.74339 & -19.954235 & 48.4 & 0.3 & 6.6 &  & 6.8 &  &  & 0.48 &   J212657-195713 & 77.6 &  &  & 0.50 &   \\
		J212708.0-204105 & 321.783572 & -20.68474 & 154.7 & 0.4 & 24.9 & 56.4 & 7.2 & 0.88 & 0.07 &  &    &  &  &  &  &   \\
		J212719.5-195353 & 321.831645 & -19.898178 & 40.6 & 0.4 & 6.1 &  & 7.0 &  &  & 0.51 &   J212719-195358 & 888.1 &  &  & 0.00 &   \\
		J212720.3-211004 & 321.834981 & -21.167881 & 40.6 & 0.3 & 7.1 &  & 7.5 &  &  & 0.39 &   J212720-211004 & 356 & 0.20 & 0.04 &  &   \\
		J212730.3-211753 & 321.876543 & -21.298258 & 28.3 & 0.4 & 5.8 &  & 7.6 &  &  & 0.52 &    &  &  &  &  &   \\
		J212739.1-184403 & 321.913103 & -18.734181 & 138.6 & 0.5 & 7.4 &  & 6.6 &  &  & 0.42 &   J212739-184359 & 600 & 0.40 & 0.03 &  &   \\
		J212757.8-191858 & 321.990948 & -19.316121 & 158.2 & 0.4 & 17.1 &  & 6.9 &  &  & 0.17 &   J212758-191851 & 609.6 & 0.30 & 0.02 &  &   \\
		J212759.8-213351 & 321.999428 & -21.56429 & 205.2 & 0.3 & 42.6 & 40.2 & 7.9 & 0.53 & 0.06 &  &   J212759-213351 & 415.3 & 0.61 & 0.04 &  &   -0.49\\
		J212800.1-191645 & 322.00074 & -19.279395 & 45.9 & 0.4 & 6.3 &  & 6.8 &  &  & 0.46 &    &  &  &  &  &   \\
		J212800.5-204156 & 322.00232 & -20.699102 & 77.9 & 0.5 & 15.2 &  & 7.4 &  &  & 0.17 &   J212800-204156 & 603.8 &  &  & 0.00 &   \\
		J212805.5-232944 & 322.022934 & -23.495761 & 1093.7 & 1.0 & 49.4 & 48.2 & 9.1 & 0.98 & 0.06 &  &   J212805-232944 & 3228.9 & 1.04 & 0.06 &  &   -0.67\\
		J212809.6-185419 & 322.040115 & -18.905469 & 125 & 0.5 & 10.8 &  & 6.7 &  &  & 0.28 &   J212809-185421 & 800.9 &  &  & 0.01 &   \\
		J212814.6-212050 & 322.061167 & -21.347329 & 36.7 & 0.3 & 7.8 &  & 7.8 &  &  & 0.39 &   J212814-212053 & 165.3 &  &  & 0.29 &   \\
		J212817.3-200525 & 322.072379 & -20.090467 & 20.8 & 0.4 & 5.5 &  & 7.2 &  &  & 0.55 &    &  &  &  &  &   \\
		
		\hline
		
	\end{tabular}
\end{sidewaystable}

\begin{figure}[hbt!]
	\centering
	\includegraphics[width=1.0\linewidth]{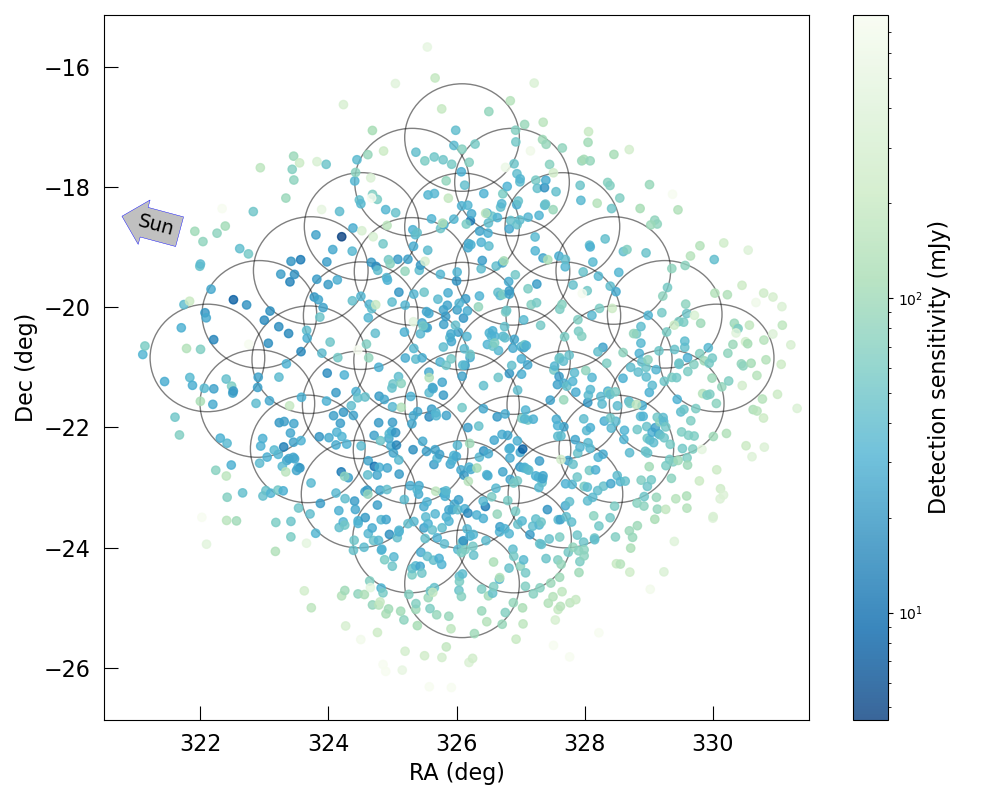}
	\caption{Sky distribution of all sources detected in our standard image. The colour axis shows the detection limit for scintillation at the position of the source. The location of the Sun is towards the upper left of the image at RA=315.3397 and Dec=-16.9462 degrees, as indicated by the arrow. Positions and size (full width of half maximum) of the 36 PAF beams are overlaid as grey circles.}
	\label{Fig:sensitivity_dist}
\end{figure}

\begin{figure}[hbt!]
	\centering
	\includegraphics[width=1.0\linewidth]{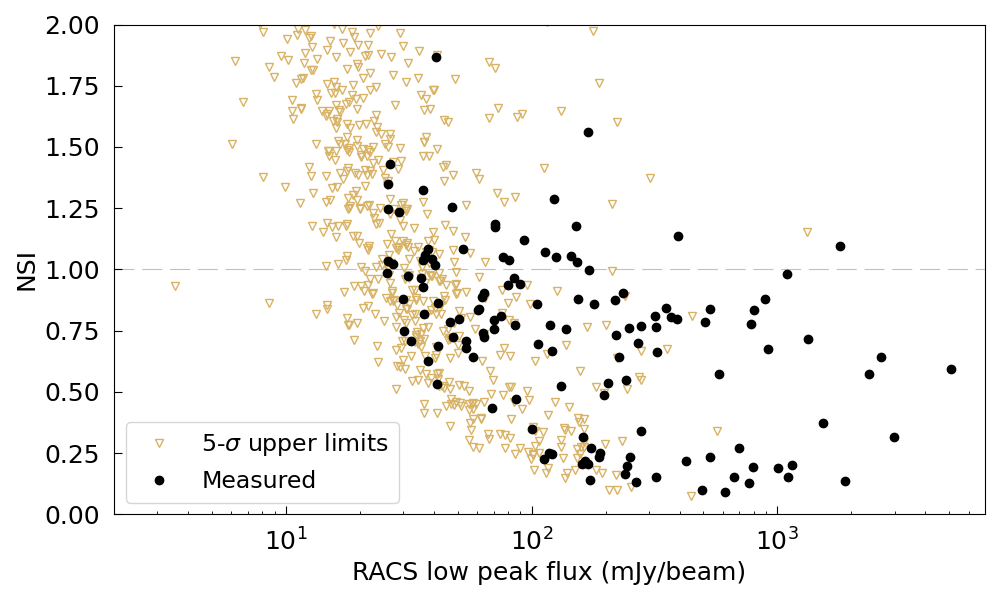}
	\caption{Distribution of NSIs and upper limits as a function of flux density in RACS low for all sources detected in our standard image.}
	\label{Fig:RACS_pkFlux_NSI-COMBINED}
\end{figure}

The sky distribution of all sources detected in the standard images  is shown in Figure \ref{Fig:sensitivity_dist}, with the colour axis indicating the the 5-sigma limit to IPS detection. The distribution of all sources detected in our standard image (both with measured NSI and upper limits) as a function of peak flux density in RACS-low is shown in Figure~\ref{Fig:RACS_pkFlux_NSI-COMBINED}. The plot shows that our observations are complete for compact components above 150\,mJy/beam down to NSI$\sim$0.1 --- i.e. we detect all compact components with flux density $>$15\,mJy. 
Additionally, we note that above $\sim$500\,mJy all objects show IPS detections (the one object at $\sim$1\,Jy which is not detected is due to it being in the outer part of the related PAF primary beam). We inspected the power spectrum of time series for sources above 500\,mJy and confirmed that they do indeed show the power spectrum shape characteristics of scintillation (ruling out spurious variability). Therefore, we find that all sources above 500\,mJy, in this observation, appear to either be smaller than or have a compact structure at $\,$0.1\arcsec\ associated with them.

\subsection{Separation of two populations}

The measured NSI values, or their upper limits, are plotted against peak flux density in Figure~\ref{Fig:RACS_pkFlux_NSI-COMBINED}. There is a clear sensitivity limit boundary, but it is not completely sharp as individual sources are at different distances from their beam centre and have different solar elongations. 
Figures~\ref{Fig:SNcont_NSI_COMBINED} and \ref{Fig:RACS_pkFlux_NSI-COMBINED} show a gap between the high and low NSI values, and  the histogram of NSI values for all sources with detected compact structure shown in the top panel of Figure~\ref{Histo:underlying_pop-b_approach3} has a bimodal distribution with peaks at NSI values of 0.2 and 0.8.  This indicates that we are probing two distinct compact radio component populations. 

\begin{figure}[hbt!]
	\centering
	\includegraphics[width=0.80\linewidth]{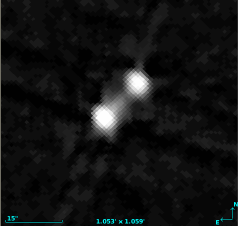}
	\caption{An example of low NSI source in the high resolution VLASS image, clearly showing two lobes. The source is J213805.0-184330 with ASKAP NSI=0.14.}
	\label{Fig:FR-II}
\end{figure}

To investigate this further, we inspected the scintillating sources with flux densities greater than 500\,mJy using images from RACS-low, RACS-mid and the VLA Sky Survey \citep[VLASS;][]{Lacy2020}. The VLASS ``quick look images'' are part of the ongoing survey made at 3\,GHz using the Jansky Very Large Array with an angular resolution of $\sim$2.5\arcsec. Although these high angular resolution images are made at much higher frequency than our IPS observations, we found that the sources with NSI$<$0.4 are mostly bright FR-II type sources \citep[][]{Fanaroff_Riley_1974}, with hotspot(s) in their lobe(s). While the cores were also visible in some cases in the sensitive VLASS images, in most cases these compact cores would have been below the detection threshold of our observation. An example of a low NSI source in VLASS quicklook median stack image obtained using Aladin \citep{2000A&AS..143...33B} is shown in Figure \ref{Fig:FR-II}. We conclude that it is the hot-spots in the lobes of these large radio galaxies that are detected as the scintillating components in the radio galaxies in the low-NSI population.  

All sources with NSI$>$0.4, were found to have a single component in the VLASS image. Sources with an NSI in the lower range of this subset showed hints of extended structure but those with higher NSI appeared unresolved. In the searches made using NASA extragalactic database (NED) many of these sources were either classified as quasars or had a generic ``radio source'' classification.  We conclude that these high NSI sources have a dominant central core which is the scintillating component. A number of these sources have Very Long Baseline Interferometry (VLBI) images\footnote{https://astrogeo.org/} with clear detections revealing a core-jet morphology. These VLBI detections are at angular scales an order of magnitude or two finer than those probed by our IPS observations. so are consistent with our high NSI values. 

We discuss the spectral properties and host identifications of sources with compact structure in more detail in Section~\ref{ssec:spectra}.

\subsubsection{Estimating the detected fraction of NSI}
\label{Sec:underlying_NSI}

\begin{figure}[hbt!]
	\centering
	\includegraphics[width=1.0\linewidth]{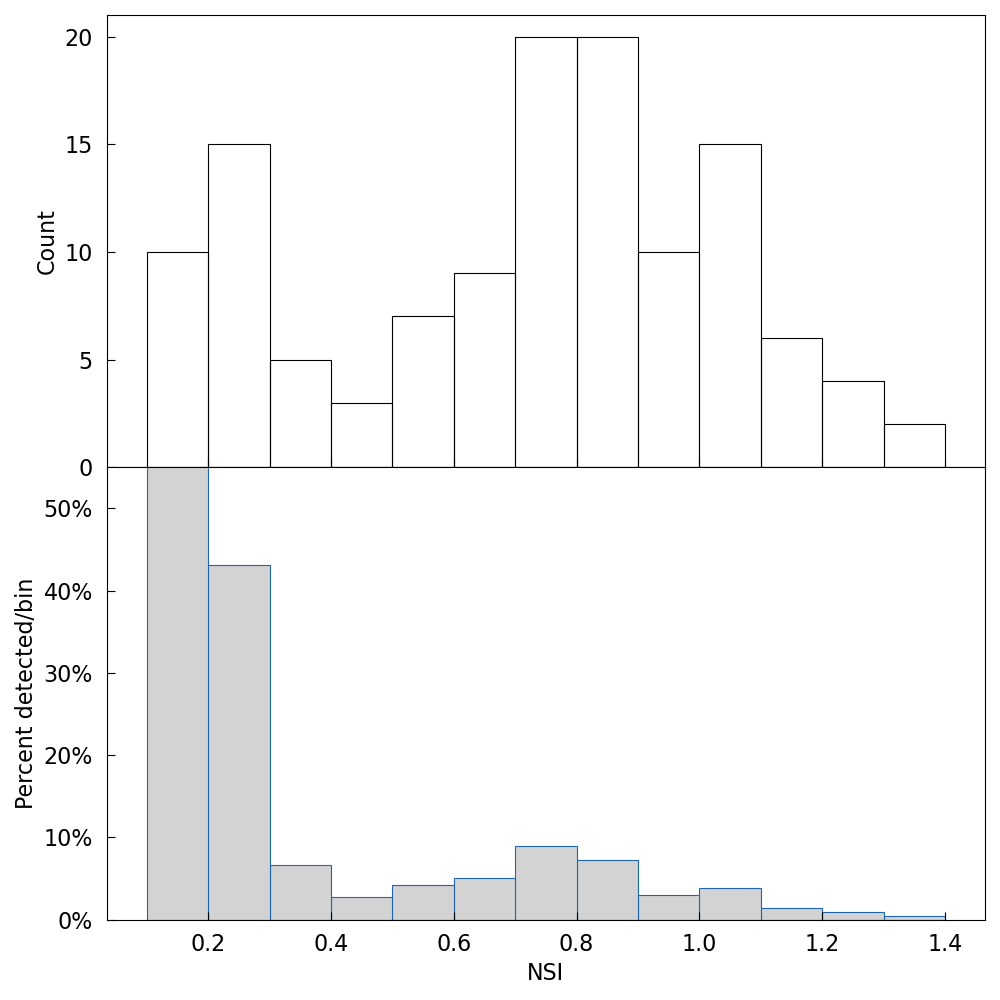}
	\caption{Histogram of the number of sources with measured ASKAP NSI in 0.1 width bins (top panel), and our estimate of the fraction of sources with detected ASKAP NSI (bottom panel). The Y-axis is truncated at 55\% to better show the distribution.}

	\label{Histo:underlying_pop-b_approach3}
\end{figure}

Next we estimate the fraction of sources for which compact components are detected in our catalogue as a function of their NSI level. 
As is clearly shown in Figure \ref{Fig:RACS_pkFlux_NSI-COMBINED}, we can detect high-NSI sources down to much fainter total flux densities than low-NSI sources. This provides a larger sample of sources at high NSI values compared to low NSI values. Therefore, direct counts of number of detected source numbers as a function of NSI level (upper panel of Figure \ref{Histo:underlying_pop-b_approach3}) is strongly biased, with artificially high counts in the higher NSI bins.
This bias can be corrected by counting the sources with measured NSIs and using the upper limits on NSI in our catalogue to normalise the detected number of sources by the total number of sources that could have been detected at each NSI. This was done for all individual detected sources before binning the NSI values. The result is presented in Table \ref{Table:underlying_NSI_Approach3}, and plotted in Figure \ref{Histo:underlying_pop-b_approach3} lower panel.
Note that these are the fraction of NSI values in each bin.  If we sum over all NSI bins $>$0.40, 38\% of all bright ASKAP sources at 823\,MHz have significant compact components, and 17\% with NSI $>$0.8 are consistent with being single point sources.  

\begin{table}[hbt!]
	
	\caption {Estimate of the fraction of sources for which NSI is detected.}
	\label{Table:underlying_NSI_Approach3}
	\begin{tabular}{llcr}
		\toprule

		  NSI Bin &  & Detected   &  \%       \\ 
		low & hi  & in bin     & detected  \\
		\hline
		1.3 & 1.4 & 2 & 0.4 $\pm$ 0.3\\
		1.2 & 1.3 & 4 & 0.9 $\pm$ 0.4\\
		1.1 & 1.2 & 6 & 1.4 $\pm$ 0.6\\
		1.0 & 1.1 & 15 & 3.8 $\pm$ 1.0\\
		0.9 & 1.0 & 10 & 3.0 $\pm$ 0.9\\
		0.8 & 0.9 & 20 & 7.2 $\pm$ 1.6\\
		0.7 & 0.8 & 20 & 9.0 $\pm$ 2.0\\
		0.6 & 0.7 & 9 & 5.0 $\pm$ 1.7\\
		0.5 & 0.6 & 7 & 4.2 $\pm$ 1.6\\
		0.4 & 0.5 & 3 & 2.7 $\pm$ 1.5\\
		0.3 & 0.4 & 5 & 6.6 $\pm$ 2.9\\
		0.2 & 0.3 & 15 & 43.1 $\pm$ 11.1\\
		0.1 & 0.2 & 10 & 87.8 $\pm$ 27.8\\
		
		\hline
		
		\bottomrule
	\end{tabular}
\end{table}

\subsection{Comparing NSI measurements between MWA and ASKAP}
Comparing the NSIs that we have measured at 823\,MHz against NSIs measured at lower frequencies help us to understand how the population of sources with compact structures change as a function of frequency. The observations presented in this paper are also covered by the IPS survey made using the MWA at 162\,MHz \citep{Waszewski2025_PhD, Morgan2022_IPS_DR1}. In order to compare the NSIs measured at these two frequencies, we made a positional crossmatch between the RACS-low catalogue positions of our sources against the sources in the MWA IPS catalogue using a 60\arcsec\ search radius. Like the catalogue presented here, the MWA catalogue also includes upper limits. Of all sources in out catalogue, 507 sources have MWA IPS counterparts. Of these 133 sources have reliable detections in the MWA IPS catalogue and 96 of these sources have NSI detections in both ASKAP and MWA IPS catalogues, and we compare the NSI of the sources in common in Figure~\ref{Fig:CompareNSI_ASKAP-MWA}. 

\subsection {Frequency dependence of NSI}
\begin{figure}[hbt!]
	\centering
	\includegraphics[width=1.0\linewidth]{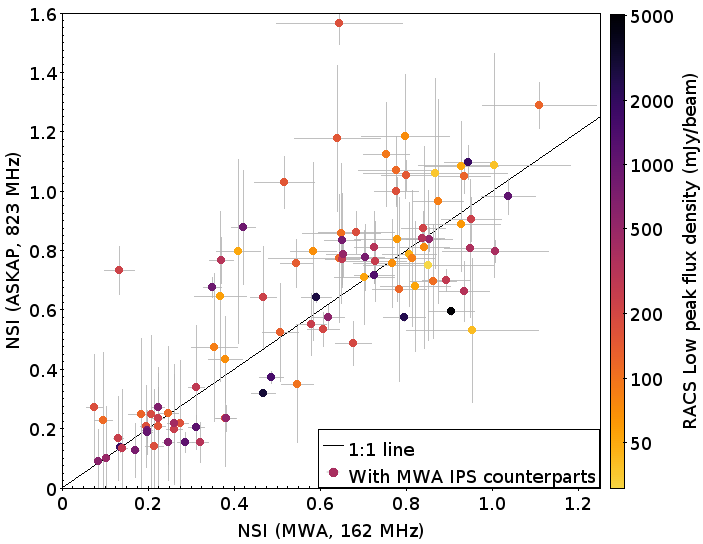}
	\caption{Comparison of ASKAP and MWA NSIs. The black line shows the 1:1 line, and the colour of the dots show their peak flux density in RACS low.}
	\label{Fig:CompareNSI_ASKAP-MWA}
\end{figure}

There is a good correlation between the NSIs calculated at 162\,MHz (MWA) and 823\,MHz (ASKAP), but with significant scatter. The ratio of NSIs between the two frequencies is a combination of the spectrum of the compact component, the changing fraction of the compact component to the total source, and the frequency-dependent angular resolution of IPS. This complicates the interpretation of Figure \ref{Fig:CompareNSI_ASKAP-MWA} since three quite different processes are involved. 
The extended low-brightness steep-spectrum structure of radio galaxies detected at low frequencies will be weaker and may even fall below the detection limit at the higher frequency. In such a case the compact component at the higher frequency ASKAP observation becomes a larger fraction of the total flux density. The sources above the 1:1 line in Figure~\ref{Fig:CompareNSI_ASKAP-MWA} are mostly due to this effect.
Secondly, the relevant angular scale for IPS (the Fresnel scale) is wavelength dependent, and is $\sim$0.3\arcsec\ for MWA at 162\,MHz and $\sim$0.1\arcsec\ for 823\,MHz. Therefore, if the compact component is partially resolved in Fresnel scale at the high frequency, these objects will have lower NSI in our ASKAP observations. We see such a lowering of NSI in our observations for a number of sources, which form the sharp lower boundary in Figure~\ref{Fig:CompareNSI_ASKAP-MWA}.

Finally, the compact component is likely to have a different spectrum than the total source and we investigate this in Section~\ref{ssec:spectra}.

\subsection{Spectra of sources with detected IPS}

The GLEAM catalogue provides spectral energy distribution for sources at 20 frequencies between 70\,MHz and 230\,MHz and provides spectral indices for sources that exhibit power-law behaviour within its observing band. 
For all sources in our sample, we also calculated the spectral indices of the  flux densities between the GLEAM wideband 200\,MHz (which provides the most sensitive GLEAM measurements), and the sensitive RACS-low 888\,MHz (which adds ~5\% error in flux density compared to using 823 MHz) but doesn't affect the the overall spectral behaviour significantly.  

\begin{figure}[hbt!]
	\centering
	\includegraphics[width=0.95\linewidth]{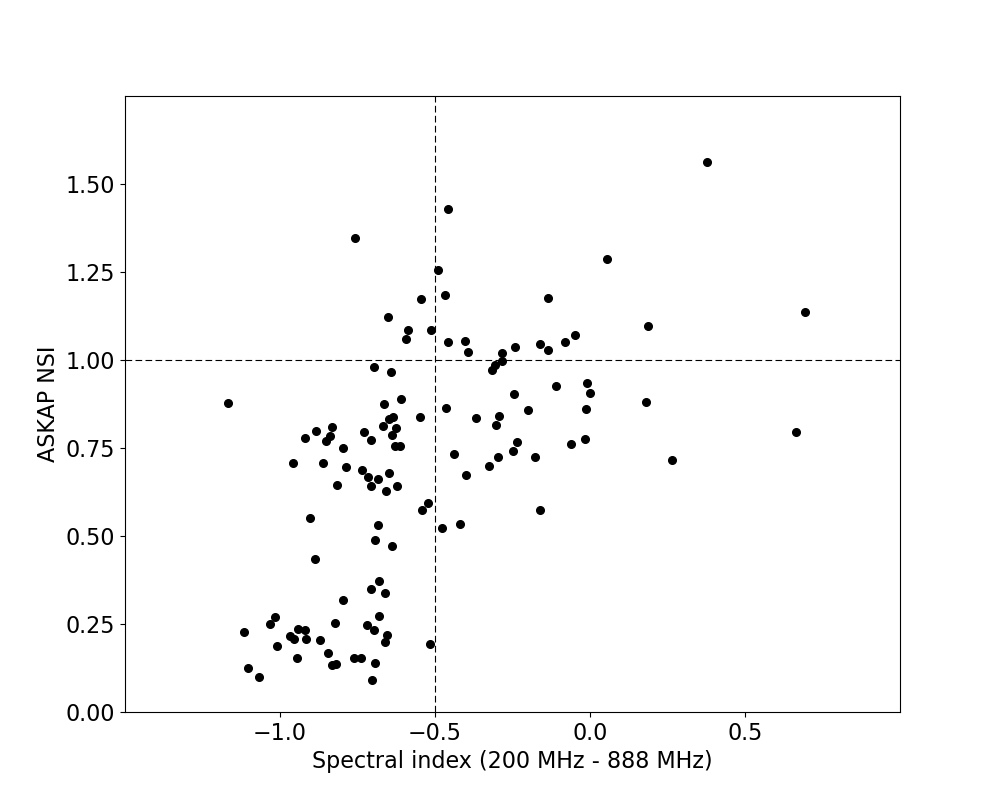}
	\caption{Spectral index distribution of sources with detected  scintillation. Spectral index of the total source has been calculated between GLEAM 200\,MHz and RACS Low 888\,MHz. Vertical dashed line has been drawn at the widely accepted boundary between flat- and steep-spectrum.}
	\label{Fig:alpha_NSI}
\end{figure}

In Figure~\ref{Fig:alpha_NSI}, we plot the distribution of NSI as a function of their spectral indices between 200\,MHz and 888\,MHz. The distribution of sources in this plot is reminiscent of the distributions at 162\,MHz of IPS sources identified with the MWA, and at 20\,GHz made using the AT20G high-angular-resolution catalogue \citep[][figures 12 and 7 respectively]{Chhetri2018_IPS2, Chhetri2013-AT20G_HARC}. 

\begin{figure}[hbt!]
	\centering
	\includegraphics[width=0.95\linewidth]{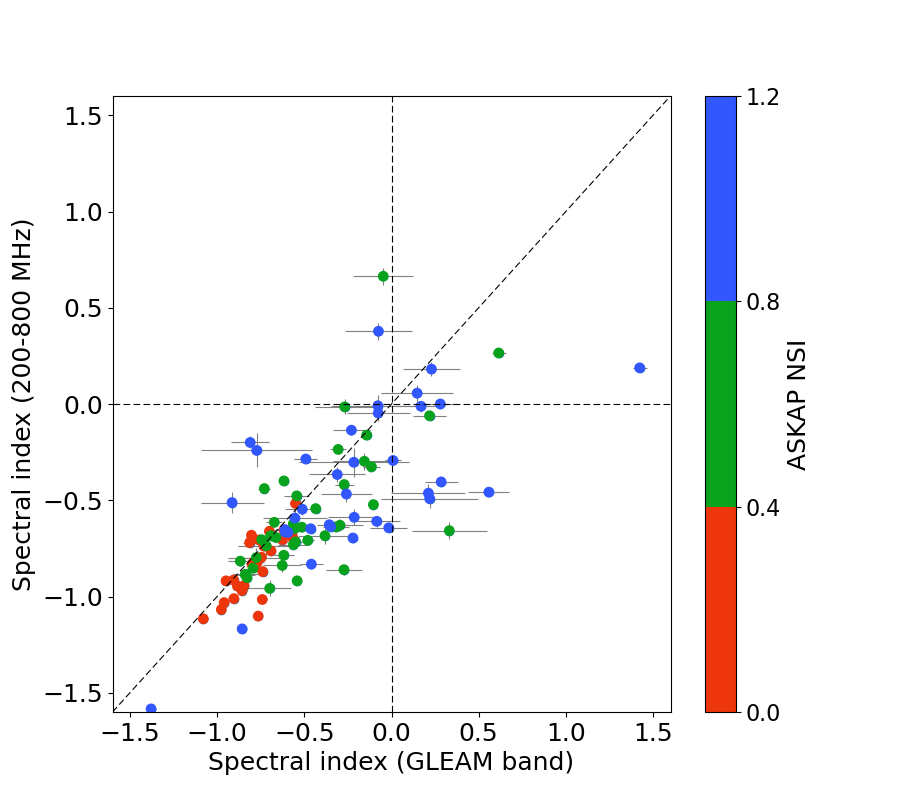}
	
	\caption{Colour-colour plot of scintillating sources. Abscissa: spectral index within the GLEAM bands. Ordinate: spectral index between 200\,MHz and 888\,MHz. Colours indicate ASKAP NSI.}
	\label{Fig:col-col_radio}
\end{figure}

In Figure~\ref{Fig:col-col_radio}, we show the colour-colour plot for these sources, with the GLEAM spectral indices in the X-axis and spectral indices between GLEAM 200\,MHz and RACS-low 888\,MHz in the Y-axis. The colour bar roughly divides the NSI into three bins: weakly scintillating sources (0$<$NSI$<$0.40), moderately scintillating sources (0.40$\geqslant$NSI$<$0.8) and strongly scintillating sources (NSI$\geqslant$0.8). We now explore these three subgroups. 

As noted in the previous section, the weakly scintillating sources are the objects that were found to be dominated by hot-spots in their extended components (lobes). These sources, represented with red coloured dots in Figure \ref{Fig:col-col_radio}, clearly lie along the 1:1 diagonal line representing power law (with index between $\sim$-0.5 and -1.1). The lack of scatter about this line by these objects means that they maintain their power-law trend across a wide frequency span, ranging from 70\,MHz to 888\,MHz. These are the FRII sources.

The moderately scintillating sources, shown with green coloured dots in the Figure, have more scatter both above and below the power-law line, but with higher fraction of objects below this line indicating that the spectra of the majority of moderately scintillating objects become flatter with higher frequencies. 
These could be compact core-jet structures with a flatter spectrum core and a steeper spectrum jet which is partially resolved by the IPS.

Finally, the strongly scintillating sources, shown with blue coloured dots, show the highest scatter in spectra in Figure~\ref{Fig:col-col_radio}. A large fraction of objects lie below the power-law line indicating that the majority of the objects have spectra which have turned over and become steep-spectrum at higher frequencies. Since, we expect little or no extended components associated with these high NSI sources, we can be confident that the change in spectra is from the scintillating component, telling us that these are peaked-spectrum objects that show peak between the frequencies of 70\,MHz and 888\,MHz. A small number of outliers exist in this population which are very compact but have steep low frequency spectra flattening at higher frequencies. These are pobably CSS sources with flat-spectrum core but further study is needed.  

 Gordon-Hall et al (2026, PASA accepted) have made a detailed analysis of the integrated radio SEDs for a bright source subsample from this catalogue.  They find 47\% of sources with NSI $>$ 0.8 are peaked spectrum while no sources with NSI $<$ 0.4 have a peaked spectrum.  This is consistent with $\sim$40\% in \cite{ODea1998}) but larger than the fraction used in many models (e.g. $<$25\% in \cite{Jackson_1999MNRAS.304..160J}).
A small fraction of very high NSI sources (9\%) have flat or inverted spectrum typical of a blazar population. With future IPS observations with ASKAP, we will be able to greatly enlarge the sample of sources from a wider area of the sky to investigate source populations in detail.

\subsection{Spectra of compact objects}
\label{ssec:spectra}
We now focus our attention on the spectra of the compact components in these radio sources. We estimate the flux density that arises from just the compact component by multiplying the flux density of the total source
by its NSI. Using these flux densities at ASKAP and MWA frequencies, we calculated the spectral indices of the compact component, and these are presented in Table \ref{Table:catalogue}. We note that the compact objects we are discussing here can either be a single compact source with NSI $\sim$1, or compact components (e.g. cores or hot-spots) that are embedded within extended structures. The MWA flux densities at 162\,MHz were produced by taking the average over the GLEAM subbands centred at 158 and 166\,MHz, which exactly match the bandwidth used for MWA IPS observations.

\begin{figure}[hbt!]
	\centering
	\includegraphics[width=0.95\linewidth]{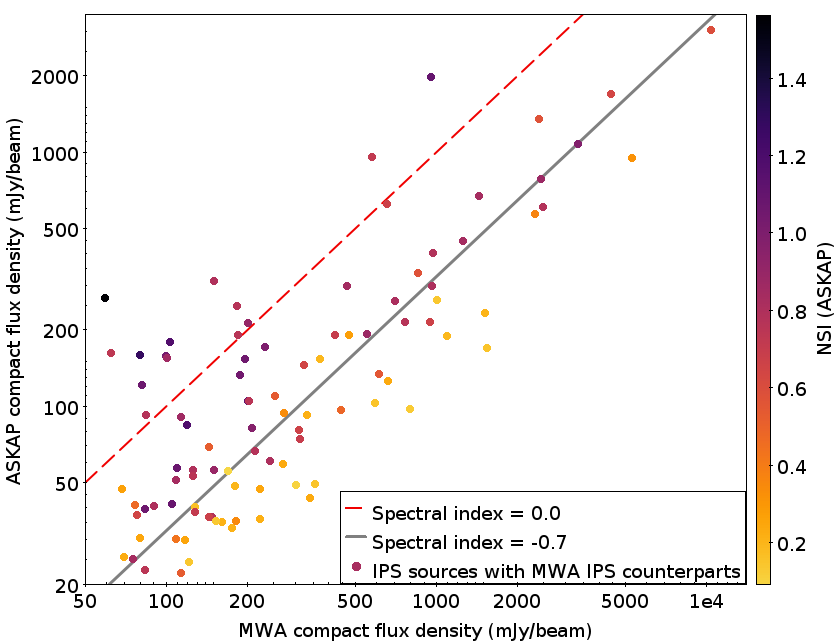}
	
	\caption{Comparison of compact component flux density for the sources in MWA at 162\,MHz (X-axis) and ASKAP at 823\,MHz (Y-axis). The red dashed line and the grey line show the flux density trend for spectral indices of 0.0	 and -0.7 respectively. The colour of the dots show ASKAP NSI for the sources.}
	\label{Fig:CompareCompFlux_ASKAP-MWA}
\end{figure}

In Figure~\ref{Fig:CompareCompFlux_ASKAP-MWA}, we compare the flux density of the compact component as measured by the MWA and ASKAP providing a different perspective to that seen in  Figure~\ref{Fig:CompareNSI_ASKAP-MWA}. In this plot we see a large number of components with a steep power-law spectrum $\alpha$ = -0.7. The lowest NSI sources have components with an even steeper apparent spectrum, but this may be a resolution effect due to the wavelength dependent Fresnel scale discussed previously. There is also a population with relatively high NSI and a wide range of spectra distributed around spectral index $\alpha$=0.0.  This is what we would expect for a two point spectral index if these compact components have a spectrum peaking anywhere between 162\,MHz and 823\,MHz.
We took the subset of low NSI sources following the power-law trend and estimated the spectral indices of their compact components between 162 and 823\,MHz. The mean value of their spectral indices is $\alpha$ = -0.73. This spectral index value is characteristic of hot-spots in lobes which once again matches with our finding in the previous section that the compact components in low NSI sources are hot-spots.

\begin{figure}[hbt!]
	\centering
	\includegraphics[width=1.0\linewidth]{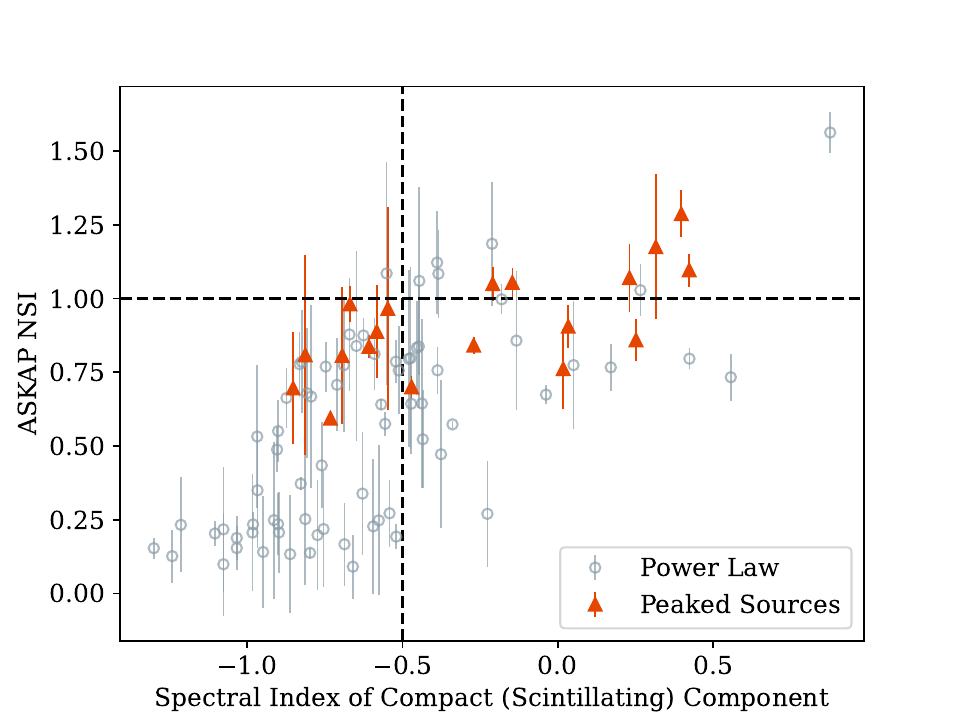}	
	\caption{Distribution of spectral index of the compact component with ASKAP NSI. Spectral index has been calculated using flux densities arising from the compact component as discussed in the text.}
	
	\label{Fig:alpha_NSI_Compact}
\end{figure}

In Figure~\ref{Fig:alpha_NSI_Compact}, presented in Gordon-Hall et al (2026, PASA accepted) we present the distribution of ASKAP NSI as a function of the spectral index of the compact component. A significant difference is noticeable between this plot, and its counterpart using the spectral index of the total sources shown in Figure~\ref{Fig:alpha_NSI}. At the low NSI part of Figure~\ref{Fig:alpha_NSI}, the compact component contributes less compared to the extended component, therefore the total spectrum is dominated by the steep-spectrum extended component. Upon removing this contribution of the extended component, it is revealed in  Figure~\ref{Fig:alpha_NSI_Compact} that the spectra of the compact objects at low NSI is still steep, but it is now more widely distributed (ranging from -0.5 to -1.25). The steepness is consistent with our above finding that the low NSI population is composed of hot-spots, but with a broader range of spectra, due to the complexity of the hotpots. This may be related to the energy supply from the core, and may indicate their different stages (e.g. the most steep-spectrum one indicative of a hot-spot of a radio galaxy before the remnant stage, and the flatter-spectrum indicative of the turnover in hot-spot spectra as shown by \cite{McKean2016} for Cygnus-A). 

The large fraction of sources with NSI in the range between 0.6 and 0.85 appear to have extended towards flatter spectrum compared to Figure~\ref{Fig:alpha_NSI}. They include many peaked spectrum sources and the spread in spectra is a result of the fixed frequency range used which can be either side of the peak. But if they are at high redshifts, they would be found with steep spectrum. Therefore, these are good candidates for high redshift radio sources.  Another interesting possibility is that sources with two equally scintillating components would produce a NSI of 0.7.

High frequency peakers (e.g. gigahertz peaked-spectrum sources) and blazars, as well compact symmetric objects (CSOs) occupy the NSI$\sim$1 band. The latter two population may be distinguished using multi-epoch variability studies, and their mid-infrared colours (Chhetri et al. in prep).

We note that the outlier source in Figure~\ref{Fig:alpha_NSI_Compact}  with apparent NSI of 1.6 and spectral index of +1 is J215031.1-223200 a known highly variable blazar. Since the ASKAP and MWA observations were taken at different epochs, this variability may be the reason for the very positive apparent spectral index of the compact component of this source. This source and J214516.0-224029 also have unusually high NSI values that cant be easily explained as just IPS.  They may have additional very short period variability and warrant further investigations. 

\subsection{Extension to future ASKAP IPS observations}
As described above, these observations were part of a technical test carried out to explore the viability of IPS research with ASKAP. As part of future work, we plan to build on the results presented here via further observations that are collecting data over time on discrete fields that tile the sky within a few degrees of the ecliptic. Those fields at a suitable distances from the Sun (e.g. between 10\degr--30\degr\ in angular separation) are targeted whenever IPS observations are scheduled, optimising for sky coverage (although multiple observations per field allow us to measure, and smooth over solar wind variability). Currently, this strategy is envisaged to make use of ASKAP's autonomous observation scheduling with the goal of ``filling gaps'' around the high-priority survey science program (SSP) observations, while also making use of the under-subscribed daytime hours which are less preferred by some surveys due to the impact on survey data of the Sun in the distant sidelobes. In the long term, we hope to transition this mode of observations to one that can be carried out regularly with automatically-processed data, in order to minimise the impact on CRACO disk space and maximise our ability to capitalise on available ASKAP time. Such multiple observations will provide a robust baseline scintillation level for sources observed, and is one avenue to detect transient astronomical phenomenon between 0.1\,second and $\sim$3 minute time scales as well as to monitor the solar wind and detect transient space weather events.

\subsection{Future ASKAP IPS data processing}
We are currently adapting our data reduction pipeline to make use of modern tools more suited for wide-field imaging in a scalable manner on high performance computers. This work is being conducted under a project by Astronomy Data and Computing Services (ADACS) through it's merit allocation program. The resulting pipeline will allow us to process data using high performance computers in a timely and automated manner.

\section{Summary}
We presented the first IPS observations made with ASKAP using all 36 PAF beams, covering a field of view of 35 square degrees and detecting 131 scintillating sources in a single 2.5 minute-long observation. With these observations we demonstrate the usability of ASKAP to obtain high quality IPS detections at 823\,MHz.

There is a clear bimodality in the distribution of Normalised Scintillation Indices. A cut made at an NSI of $\sim$0.45 cleanly separates sources that are core-dominated (NSI$>$0.45), and sources that we have identified as having emission from compact hot-spots in their lobes (NSI$<$0.45).  This may provide a simple classification of the FR-II class of radio source.  We estimate that 38\% of the sources detected in a 823\,MHz survey have significant compact structure (NSI > 0.4) and that 19\% are totally core-dominated (NSI > 0.8). A large fraction of these compact sources have peaked spectra.

 We have used the MWA IPS survey at 162\,MHz to estimate the spectral indices of the scintillating compact components.  The-core dominated sources have a very wide range of spectra which is a mixture of peaked spectrum sources and blazars with more complex and variable spectra.  The hot-spots have a wider range of spectral indices than the extended structure (ranging from -0.5 to -1.25) , with the majority having spectral indices $<$-0.5.

We calculate a density of scintillating sources in our ASKAP survey down to 15\,mJy at 823\,MHz of 3.7 sources/deg$^2$. This is ideal for monitoring space weather structure.  These results only used a partial ASKAP bandwidth (120\,MHz). With recent upgrades to CRACO, it will be possible to use the full 288\,MHz bandwidth currently implemented in the ASKAP telescope, improving the sensitivity of IPS observations with ASKAP, and thus, the density of detected sources in future studies.

All these results are based on 2.5 minutes of ASKAP observations, making a full survey of the ecliptic feasible with a realistic amount of observing time. To this end, we are continuing to observe a growing number of fields with ASKAP, and are putting resources into developing a robust and scalable data reduction pipeline.

\section*{Data Availability}
Catalogue related to this article is submitted along with this
manuscript. Upon acceptance, the catalogue will be available
through VizieR. Any further information will be provided
upon request to the authors.

\begin{acknowledgement}
	This research has made use of the NASA/IPAC Extragalactic Database (NED),
	which is operated by the Jet Propulsion Laboratory, California Institute of Technology, under contract with the National Aeronautics and Space Administration.
	This research has made use of "Aladin sky atlas" developed at CDS, Strasbourg Observatory, France
	
\end{acknowledgement}
\bibliography{ref_ASKAP_IPS}

\appendix

\end{document}